\documentclass[apj]{emulateapj}

\usepackage[usenames,dvipsnames]{color}
\usepackage{multirow}
 \newcommand{\um}{$\mu$m }

\begin{document}

\title{Compositional Similarities and Distinctions \\
between Titan's Evaporitic Terrains}

\author{S. M. MacKenzie\altaffilmark{1}}
\affil{}

\email{mack3108@vandals.uidaho.edu}
\and
\author{Jason W. Barnes\altaffilmark{1}}
\affil{$^{1}$Department of Physics, University of Idaho, Moscow, ID 83844-0903 USA}

\begin{abstract}
We document the similarities in composition between the equatorial basins Tui Regio, Hotei Regio, and other 5-$\mu$m-bright materials, notably the north polar evaporites, by investigating the presence and extent of an absorption feature at 4.92 $\mu$m. In most observations, Woytchugga Lacuna, Ontario Lacus, MacKay Lacus, deposits near Fensal, some of the lakes and dry lake beds south of Ligeia, and the southern shores of Kraken Mare share the absorption feature at 4.92 \um observed in the spectra of Tui and Hotei. Besides Woytchugga and at Fensal, these 5-$\mu$m-bright deposits are geomorphologically-substantiated evaporites. Thus, the similarity in composition strengthens the hypothesis that Tui and Hotei once contained liquid. Other evaporite deposits, however, do not show the 4.92 \um absorption, notably Muggel Lacus and the shores of Ligeia Mare at the north pole. This difference in composition suggests that there are more than one kind of soluble material in Titan's lakes that can create evaporite and/or that the surface properties at the VIMS wavelength scale are not uniform between the different deposits (crystal size, abundance, etc). Our results indicate that the surface structure, composition, and formation history of Titan's evaporites may be at least as dynamic and complex as their Earth counterparts.

\end{abstract}


\section{Introduction}
Titan, Saturn's largest moon, is swathed in a thick atmosphere with haze that is highly scattering and gas that absorbs at visible wavelengths. Several wavelengths of light in the near infrared penetrate the atmosphere to reach the surface, however, with varying degrees of interference from the atmosphere \citep[e.g.][]{1991Icar...93..362G,2003Icar..162..125L,2004GeoRL..3117S02G}. Since \emph{Cassini}'s arrival at the Saturnian system in 2004, the Visual and Infrared Mapping Spectrometer (VIMS) has been observing Titan's surface at these seven spectral windows where methane gas absorption is weakest in the IR \citep{2004SSRv..115..111B,2005EM&P...96..119B,2005Natur.435..786S}. The visible channels can be also be used to look at the surface, but to a more limited extent \citep{Vixie2012}.\\

Even through these windows, however, observing the surface is complicated by the properties of Titan's atmosphere. The haze particles of are highly forward scattering \citep[e.g.][]{2004SSRv..115..363P,2008P&SS...56..669T}. This effect is seen across the IR-spectrum, increasing the reflectance of both windows and methane-saturated bands, and is stronger at the shorter wavelengths than at longer ones. Analyses of solar occultations \citep{2014Icar..243..158H} and specular reflections \citep{2013SpecRef} have measured the decreasing total optical depth (both haze and gas opacities) as a function of increasing wavelength. 
The haze distribution changes with latitude \citep[e.g.][]{2005Sci...308..975F,2010Icar..208..850R}, altitude \citep[e.g.][]{2002AJ....123.3473Y}, and season \citep[e.g.][]{2001Icar..152..384L, 2015Icar..250...95V}. Scattering properties of the haze particles change with altitude and wavelength \citep{2005Natur.438..765T,2008P&SS...56..669T}. Despite this wealth of new insight into Titan's atmosphere in the \emph{Cassini} era, the details of the atmosphere's effects on emergent spectra are still far from understood. \\

Thus, as VIMS IR data only sample seven parts of Titan's surface spectrum, it is difficult to identify which specific chemical species are present to create the observed signal. And yet, there are three cases where this has been done. \citet{2010JGRE..11510005C} identified a feature in Titan's global spectrum at 5.05 $\mu$m, which the authors suggest is indicative of benzene. \citet{T38.ethane} identified ethane in the VIMS data from Ontario Lacus. \\

\citet{McCord2008} identified an absorption feature in the spectrum of Tui Regio at 4.92 \um using some of the earliest VIMS data. Their method used deviance from the scene average as the criteria for absorption. CO$_2$ ice was originally proposed as a candidate material, as it demonstrates an absorption feature near 4.92 $\mu$m. \citet{2010JGRE..11510005C}, however, point out that the wavelength shift necessary to match CO$_2$ ice to the 4.92 \um feature is unphysical and the expected CO$_2$ feature in the 2.7 \um region is not observed. \citet{2010JGRE..11510005C} also list several species from their laboratory survey with spectral features near 4.92, including a predominately potassium ferricyanide mixture, acrylonitrile, and cyanonaphthalene. However, as we cannot yet even specify exactly \emph{where} this 4.92 $\mu$m absorption occurs, let alone consider other coincident spectral features, the specific attribution of a chemical species to the 4.92 \um absorption has not yet been made.\\

Generally, however, VIMS data are used to classify surface material into spectral units, groups of material that share overall spectral characteristics. Spectral units define the unique characteristics of the ``dark brown" dunes \citep{2007mapping, 2007P&SS...55.2025S,2014Icar..230..168R}, ``dark blue" terrains including mountains \citep{Rodriguez.landingsite,2007mapping,BarnesMountains,2008JGRE..113.4003L,2014Icar..230..168R}, and the mountainous, channel-ridden Xanadu region \citep{2007mapping}. \\

Titan's lakes and seas are also spectrally distinct: in the near infrared, they are darker than their surroundings at all wavelengths \citep{T38.ethane,2012Icar..221..768S}. These liquid bodies seem to be restricted to the polar regions, though see \citet{2012Natur.486..237G} and \citet{2015Icar..257..313V}. At the north pole, the lakes and seas are of a variety of shapes and sizes: from Kraken Mare, the largest sea covering 400,000 km$^2$ \citep{2009GeoRL..3602204T}, to the Lake District, a region of ``cookie-cutter lakes"\citep{Hayes2008} which are just resolvable in VIMS data. The south pole, however, looks quite different. The south has just one substantial body of liquid, Ontario Lacus, which is about 15,000 km$^2$ in size \citep{Hayes2008}. \\

Thermodynamic models \citep[e.g.][]{2009ApJ...707L.128C,2012P&SS...61...99C,2013GeCoA.115..217G} and laboratory experiments \citep[e.g.][]{2014P&SS...99...28D,2015GeoRL..42.1340M,2015P&SS..109..149H,2015E&PSL.410...75L} that explore the possible bulk composition of the lakes, as well as observations with the RADAR instrument that constrain some of the liquid's properties \citep[e.g.][]{Hayes2010,2012Icar..221..960V,Hofgartner2014}. Besides the ethane detection of \citet{T38.ethane}, VIMS data is difficult to use to confirm the bulk composition of the liquid as the methane in the atmosphere obfuscates detection of liquid methane on the surface and VIMS can only sample a depth of a few microns. The ethane absorption in Ontario Lacus's spectrum is not sensitive to abundance. No other absorption features have been observed in VIMS lake spectra. \\

Some of the lakes and seas on Titan have evaporite along their shorelines, as do some dry lakebeds \citep{evaporite, MacKenzie14}. Evaporite is solid material redeposited on the surface after the liquid in which it was dissolved evaporates. On Titan, these deposits demonstrate a unique spectral behavior such that they are referred to as the 5-$\mu$m-bright spectral unit--so named because it is significantly brighter than any other surface units in the eponymous window. At 2.8 $\mu$m, this material is relatively brighter than other surface units, but at shorter wavelengths, 5-$\mu$m-bright material is similar in reflectivity to other bright units like Xanadu or the equatorial bright unit \citep{2005Sci...310...92B}.\\

 The first evaporitic deposits were identified just south of Ligeia Mare where RADAR identified small (10-200 km in diameter) lakes \citep{evaporite} and dry lake beds \citep{Hayes2008}. The same region also has lakes and dry lake beds without the 5-$\mu$m-bright signature: evidence that the bright deposits are evaporite, the material left over when liquid (presumably methane or ethane) evaporates. Evaporite only forms once the liquid is saturated with solute (on Titan, probably some kind of hydrocarbon) and if evaporation is the dominant mechanism for removing the liquid. An evaporitic deposit therefore indicates a location where liquid has ponded on the surface at some point in the past.\\

\citet{MacKenzie14} conducted a survey of then-available VIMS data to locate all deposits of this 5-$\mu$m-bright material on Titan's surface. The authors found that while more separate instances of evaporite deposits appear at the north pole, where the majority of Titan's surface liquid is now located, the largest single deposits by surface area are located in Titan's presently desert equatorial region at Tui , the  and Hotei Regio. \citet{Moore.Tui.Hotei.Lakes} interpreted lacustrine and fluvially carved features in the RADAR coverage of these two basins as indicative of Tui and Hotei being paleo seas. Observing the 4.92 \um absorption feature in the spectra of the north polar evaporites would be further evidence for this hypothesis as it would compositionally link the basins Tui and Hotei with the clearly lacustrine-associated deposits south of Ligeia. \\

In this work, we determine the degree of similarity between the compositions of the 5-$\mu$m-bright material at different latitudes by documenting all observations of the 4.92 \um absorption in spectra of Tui Regio, Hotei Regio, and the evaporite candidates of \citet{MacKenzie14}. We compare the relative depths of the absorption feature with time and flyby geometries in an effort to distinguish what controls the absorption depth. In Section \ref{Observations}, we describe the VIMS data and in Section \ref{Methods} we describe the principal component analysis (PCA) technique that we use to analyze them. Section \ref{Results} presents our results for each 5-$\mu$m-bright deposit. We discuss our findings in Section \ref{Discussion} and conclude with a summary of our work and its implications.\\

\begin{figure*}
\includegraphics[width=\textwidth]{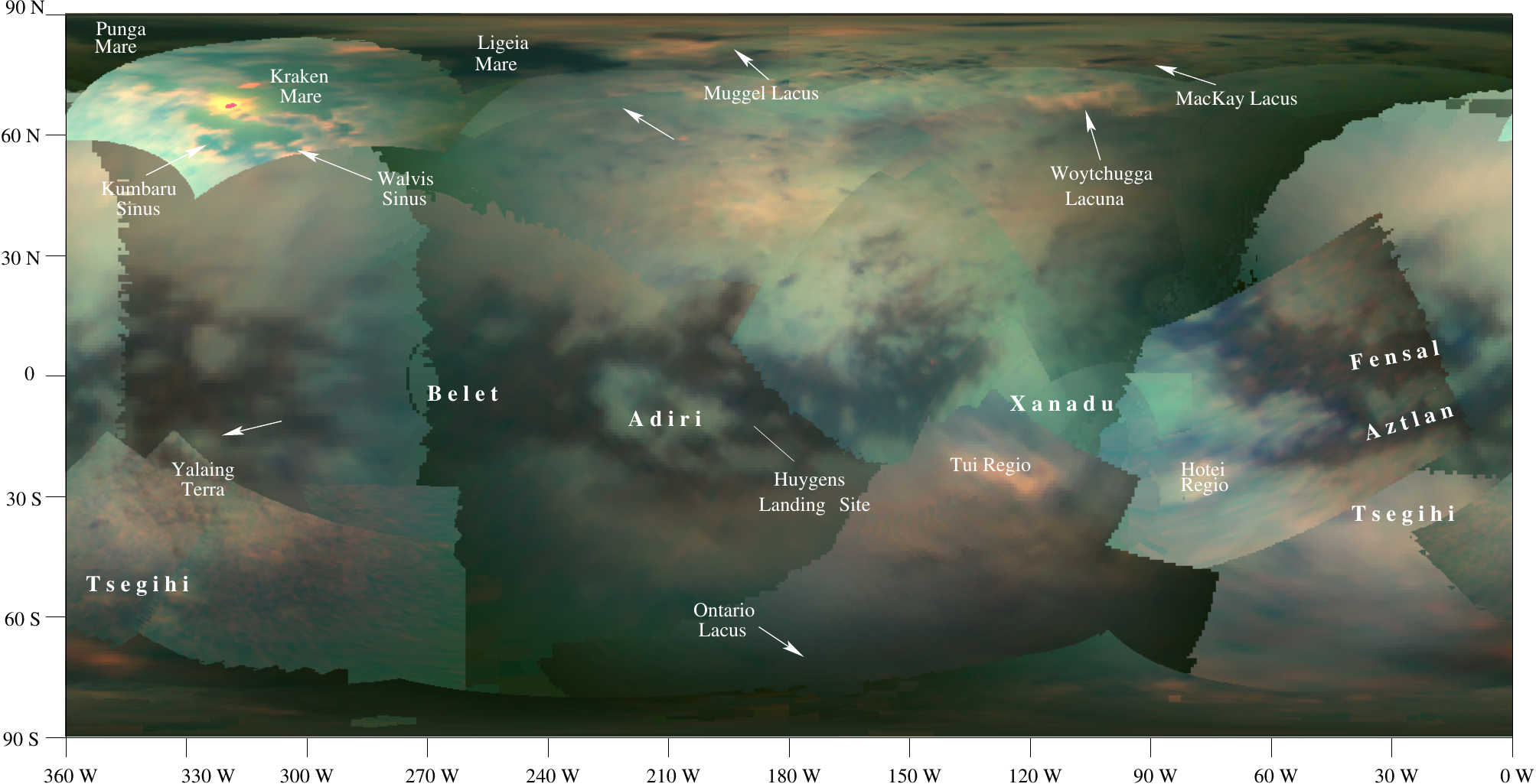}
\caption{Cylindrically projected VIMS map of Titan's surface with R = 5 $\mu$m, G = 2 $\mu$m, and B = 1.3 $\mu$m. White arrows point to the areas studied in this paper (though see Figure \ref{fig:nportho} for a better projection for the north polar deposits). Tui and Hotei Regiones are the largest 5-$\mu$m-bright deposits on Titan's surface and are easy to see at the global scale; some of the smaller evaporites are harder to distinguish in this global view.}
\label{fig:global}
\end{figure*}

\section{Observations}
\label{Observations}
We use data from the Visual Infrared and Mapping Spectrometer on-board \emph{Cassini} from 2004-2014. These were processed with the VIMS pipeline described in \citet{2007mapping} which converts the raw data into I/F, the observed reflectance divided by the incident solar flux. All geometric projections are done with software developed for \citet{T20dunes}. In Table \ref{tab::obs} we summarize the geometries and average resolutions of the data from each flyby. Tui and Hotei Regiones were imaged during 2004-2009 \citep{2009P&SS...57.1950B}, but the north pole evaporites were not fully visible until 2013-2015. The viewing geometries between equatorial and polar deposits are therefore unavoidably different. In fact, the spread of viewing geometries over just the Tui and Hotei observations differ enough as to make comparison of uncorrected I/F between flybys difficult. We describe how we address this problem in Section \ref{Methods}. \\

\begin{table*}[H]
\caption{Summary of characteristics for flybys used in this work. The incidence and emission angle ranges listed span those of the images used for the features listed. Note that shoreline evaporite deposits are listed by the name of the lake or sea they border. }
\label{tab::obs}
\begin{tabular}{c c c c c c c}
Flyby	& Date&	Features 	&	Phase	&	Inc	&	Emis	&	Best Resolution \\
	&	& Covered	&	($^{\circ}$)	&	($^{\circ}$)	&	($^{\circ}$)	&	(km/pixel)	\\
\hline
\hline
Ta	& 24-Oct-04	&Hotei, Tui, Xanadu	&	10	&	10-70	&	20-80	&	43	\\
Tb	& 13-Dec-04	&Tui, Xanadu	&	10	&	10-40	&	20-40	&	8	\\
T3	& 15-Feb-05	&Hotei, Tui, Xanadu	&	30	&	0-70	&	20-80	&	94	\\
T4	& 31-Mar-05	&Hotei, Xanadu	&	50	&	0-20	&	60	&	74	\\
T5	& 16-Apr-05	&Fensal, Hotei, Xanadu	&	50-60	&	0-40	&	20-70	&	9	\\
T6	& 22-Aug-05	&Fensal	&	50	&	40	&	20	&	40	\\
T7	& 7-Sept-05	&Fensal	&	50	&	30	&	30	&	56	\\
T8	& 28-Oct-05	&Hotei, Tui, Xanadu	&	20	&	0-50	&	20-70	&	43	\\
T9	& 26-Dec-05	&Fensal, Hotei, Xanadu	&	20	&	20-40	&	30-60	&	50	\\
T10	& 15-Jan-06	&Hotei, Tui, Xanadu	&	35	&	0-40	&	20-70	&	17	\\
T12	& 18-Mar-06	&Hotei, Tui, Xanadu	&	65	&	0-40	&	20-80	&	17	\\
T14	& 20-May-06	&Hotei, Tui, Xanadu	&	90	&	10-70	&	20-80	&	66	\\
T38	& 5-Dec-07	&Ontario	&	40	&	60	&	60	&	0.6	\\
T44	& 28-May-08	&Hotei, Tui, Xanadu	&	90	&	10-70	&	10-70	&	100	\\
T46	&  3-Nov-08	&Hotei, Tui, Xanadu	&	80	&	10-70	&	0-60	&	36	\\
T47	& 19-Nov-08	&Hotei, Tui, Xanadu	&	80	&	10-80	&	0-70	&	83	\\
T48	& 5-Dec-08	&Hotei, Tui, Xanadu	&	70	&	10-60	&	10-60	&	64	\\
T49	& 21-Dec-08	&Hotei, Tui, Xanadu	&	70	&	10-70	&	10-40	&	95	\\
T50	& 7-Feb-09	&Hotei, Tui, Xanadu	&	70	&	10-70	&	20-60	&	65	\\
T51	& 27-Mar-09	&Hotei, Ontario, Tui, Xanadu	&	70	&	10-70	&	20-60	&	39	\\
T61	& 25-Aug-09	&Kumbaru, Walvis, Yalaing	&	10	&	10-50	&	10-70	&	20	\\
T67	& 5-Apr-10	&Kumbaru, Walvis, Yalaing	&	10	&	20-50	&	10-60	&	30	\\
T69	& 5-Jun-10	&Atacama, Atitlan, Djerid, Ligeia, Uvs, Vanern	&	30	&	60-70	&	50-60	&	10	\\
T76	& 8-May-11	&South of Kraken, Walvis, Kumbaru, Yalaing	&	40	&	30-50	&	20-60	&	19	\\
T82	& 19-Feb-12	&South of Kraken, Walvis, Kumbaru, Yalaing	&	60	&	40-50	&	20-60	&	24	\\
T90	& 5-Apr-13	&Muggel, Woytchugga	&	110	&	60-70	&	40-50	&	35	\\
T93	& 26-July-13	&Muggel, Vanern,Woytchugga	&	80-90	&	50-70	&	20-50	&	7	\\
T94	& 12-Aug-13	&Ligeia, MacKay, Muggel	&	70	&	60-70	&	0-20	&	4	\\
T96	& 1-Dec-13	&Ligeia, MacKay, Muggel, Woytchugga	&	60	&	60	&	0-30	&	6	\\
T97	& 1-Jan-14	&Ligeia, Muggel, Woytchugga	&	50	&	60	&	0-20	&	13	\\
T98	& 2-Feb-14	&MacKay, Muggel, Woytchugga	&	50	&	60	&	0-20	&	74	\\
T100	& 7-Apr-14	&Ligeia, Muggel, Vanern, Woytchugga	&	30	&	50-60	&	20-40	&	19	\\
T103	& 20-Jul-14	&Kumbaru, South of Kraken, Walvis	&	100	&	40-70	&	40-50	&	95\\
T104 	& 21-Aug-14	&Flensborg, Gabes, Kumbaru,  & 100 & 40-70 & 20-60 & 29 \\
T104	& 21-Aug-14	&South of Kraken, Walvis, Woytchugga&	100&	40-70& 20-60& 29\\
\end{tabular}
\end{table*}

We investigate a select subset of the deposits identified by \citet{MacKenzie14}, including evaporites found along the shores of the seas and lakes of the north pole as well as some evaporite candidates from the equatorial regions. Figure \ref{fig:global} shows the relative locations of Tui Regio, Hotei Regio, Xanadu, and the evaporites studied here (white arrows) on a cylindrically projected VIMS map. For the coordinates and best VIMS observations of each deposit, we refer the reader to Table 1 of \citet{MacKenzie14}. \\

\section{Methods}
\label{Methods}
The 4.92 \um absorption feature shows a very shallow depth ($\sim$ 2\% difference in I/F from the scene average in Figure 27a of \citet{McCord2008}) and thus is challenging to distinguish in raw VIMS data. Although the Sun is darker at 5 $\mu$m than at shorter wavelengths, Titan's haze scatters less at this wavelength than at the shorter windows \citep{Rodriguez.landingsite}. Thus, a higher percentage of those incident photons make it to the VIMS detector unimpeded. This means that the problem in detecting the 4.92 \um absorption lies not in correcting for atmospheric interference, but in building up enough signal relative to the inherent noise (dark, read, and shot). We tried several approaches to identify the 4.92 \um absorption. In the end, the most effective method was to subtract the albedo component of the spectrum and then coadd all corrected pixels from an individual flyby that cover a certain surface feature.  \\

\subsection{Principal Component Analysis}
\label{sec:PCA}

The albedo subtraction approach that we use involves principal component analysis (PCA). In PCA, an initial spectrum $\vec{v}$, an n-dimensional vector where each entry corresponds to a VIMS wavelength channel, gets reprojected into a new space spanned by a new set of n orthogonal and linearly independent basis vectors, $\vec{b_n}$.  The new basis vectors $\vec{b_n}$ are effectively individual spectra that represent combinations of the original channels that tend to vary in concert.  The strategic purpose behind this reprojection is to simplify what can be a large $n$-dimensional spectrum into component spectra that better represent the inherent variability across a dataset.\\

PCA has a long history of use in planetary science. On the Moon, for example, PCA has been used to determine relative composition \citep[e.g.][]{1985JGR....90..805J, 2002Icar..155..285P} and to create compositional maps of similar spectral units \citep[e.g.][]{1998P&SS...46..377B, 1999JGR...10416515C}. Similarly, different spectral units and compositions have been identified on Mars using PCA \citep[e.g][]{2006JGRE..111.6007N, 2008Icar..193..112N,2010Icar..207..226G,2013Icar..225..709F}. In the outer solar system, PCA has been used to help increase the signal-to-noise ratio in Galileo data for Ganymede \citep{2008P&SS...56..406S} and VIMS data for Titan \citep{StephanePhotometricCorrection}. Recently, PCA has been used with VIMS data for Titan by \citet{SolomonidouPCA} to infer the spectral surface diversity of features in Titan's equatorial region.\\

A drawback of PCA approaches is that given the abstract and subjective nature of the results, it can be difficult to ascribe particular meaning to individual components. For instance, when we trained PCA using the full 256-channel VIMS wavelength range (0.89-5.12 $\mu$m), the resulting high-power components are non-trivial combinations of surface reflectivity, atmospheric scattering, and solar illumination. \citet{SolomonidouPCA} found similar behavior. \\

Given the concentration on the 4.92 \um feature in this work, we consider only the last 16 VIMS channels (4.84 - 5.12 $\mu$m). The total haze optical depth in Titan's atmosphere decreases strongly as a function of wavelength due to the particle size of individual aerosols ($\sim$ 1 $\mu$m) \citep{2008P&SS...56..669T,doose2015vertical}. Thus the haze influence is minimized in the 5 $\mu$m window \citep{Rodriguez.landingsite,T38.ethane}. We find that using only the sixteen 5 \um channels yields more interpretable principal components. \\

\begin{figure}
\includegraphics[width=0.5\textwidth]{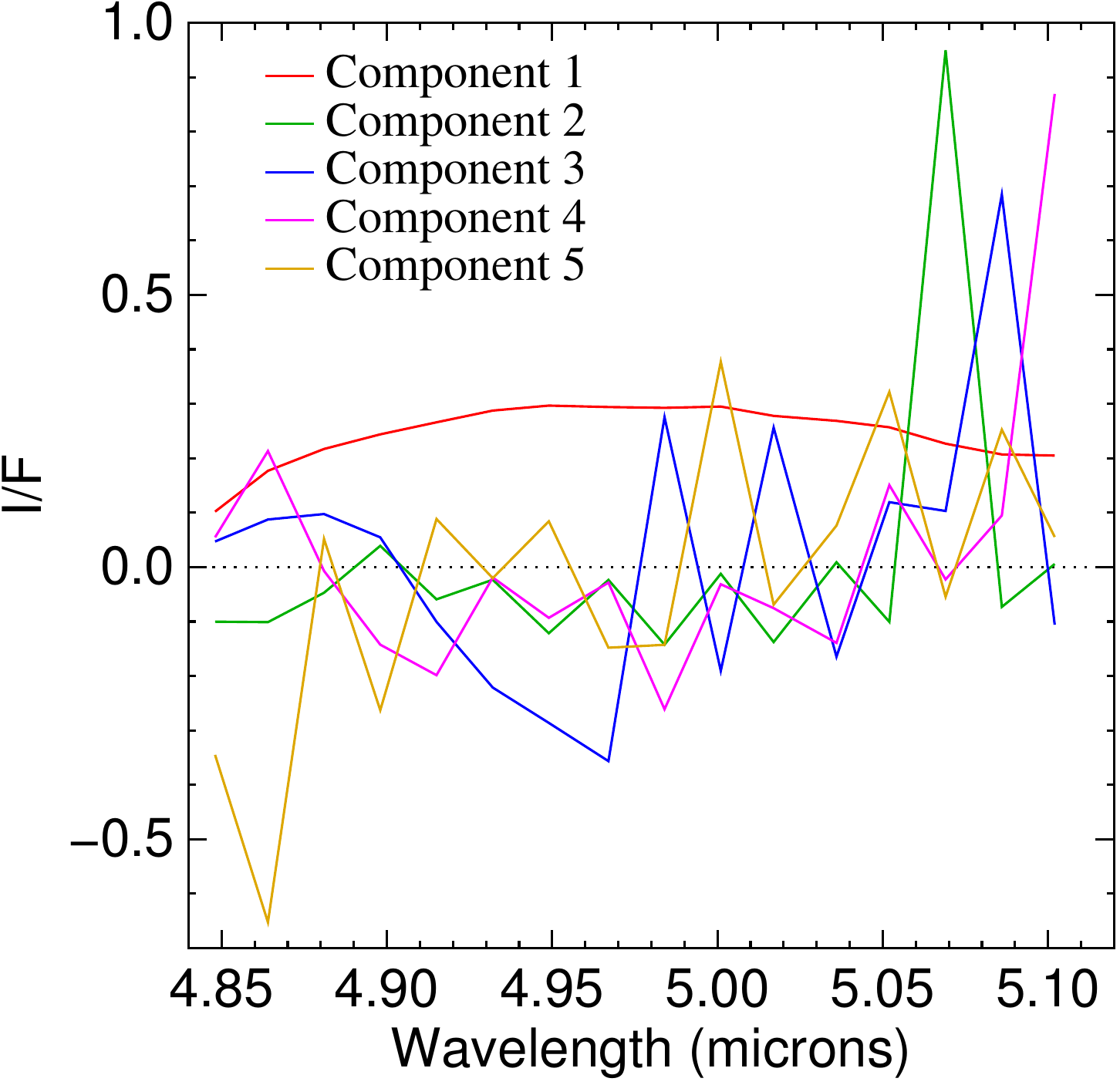}
\caption{Principal components' spectra from a principal components analysis (PCA) trained on a coadded VIMS map from T8. The first component reproduces the overall shape of the 5 \um window and is thus interpreted to represent the ``albedo" of the surface. The other components are more difficult to interpret individually, but taken together and reprojected into image space, are better for identifying shallow depth spectral features that are difficult to detect in the original VIMS spectra. }
\label{fig:components}
\end{figure}

\begin{figure*}
\includegraphics[width=\textwidth]{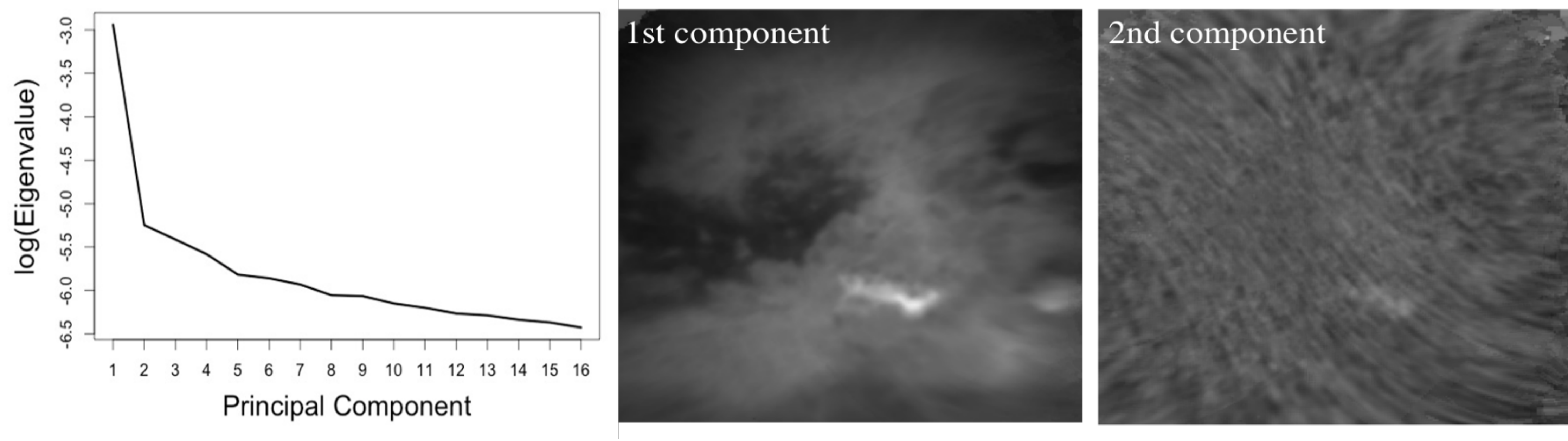}
\caption{Results of the principal component analysis for T8: (left) eigenvalues produced using the eigenvectors from T8; (middle) the first component, which we interpret to be differences in reflectance albedo; and (right) the second component. The eigenvalues describe how much variation from the mean pixel spectrum each component describes: the first component's eigenvalue is two orders of magnitude more than the next component's. Because the 4.92 \um absorption feature is so shallow, it can easily get overwhelmed by changes in the reflectance albedo due to different illumination conditions, making inter-flyby comparison difficult. Thus, to calculate the depth of the 4.92 $\mu$m absorption, we zero out off the first component from each pixel's spectrum before coadding the signal.}
\label{fig:PCAimages}
\end{figure*}

We show spectra of the first five of these components in Figure \ref{fig:components}. The shape of the first component (red line) generally matches that of the overall 5 \um window as seen in VIMS (see, for example, \citet{2007mapping}). Therefore, we interpret the first component to represent the overall surface reflectivity that, in general, varies concurrently with all channels of the 5 \um window.  We refer to this as the ``albedo" component, and  show this component as an image at the center of Figure \ref{fig:PCAimages}. The left hand graph in Figure \ref{fig:PCAimages} shows a representation of the relative magnitudes ascribed to each principal component. The first component carries over two orders of magnitude more power than subsequent principal components. \\

In the first component image (Figure \ref{fig:PCAimages}, center), the different spectral units are distinguishable. Their relative brightnesses at different wavelengths--- a function of viewing geometry, surface roughness, etc--- are the dominant control on the shape of the added pixels' cumulative spectrum. The dunes are dark while Tui, Hotei, and Xanadu are relatively bright. \\

The second component (green line of Figure \ref{fig:components}, image shown at right of Figure \ref{fig:PCAimages}) may correspond with instrumental artifacts, particularly given the high signal in the third to last VIMS channel which is known to be unreliable. Individually, however, the second component and higher orders are more inscrutable: we don't know what any of the components beyond the first truly mean. For the purpose of this analysis, interpretation of these higher-order components is in fact irrelevant.\\

If the first component represents the relative differences in albedo between different spectral units, then the remaining components should collectively be differences from the general albedo. Presumably, these differences represent compositional variation. To isolate that compositional signal, we remove the ``albedo" component to create what we refer to as ``color-only" images. \\

Functionally, we generate the color-only images by resetting the albedo component (the first principal component) to zero. We then perform an inverse principal components transformation to convert the signal back into wavelength space. In so doing, we revert back to VIMS channels ($\vec{v}$) from the more complex combinations of channels represented by the principal components ($\vec{b_n}$). \\

So in summary, we use a training dataset to infer a set of principal components. We then transform any given set of VIMS observations (the observation dataset) into the principal component space. To focus on spectral variation, we set the first component, the albedo, to zero, and then transform back into a VIMS spectrum, resulting in a color-only image cube. \\

To construct the training datasets, we coadd images from a single flyby to create a global mosaic from which we remove the signal of 5-$\mu$m-bright material. Removing the surface features of interest ensures that the first component is not controlled by the unique features of the 5-$\mu$m-bright unit. We then perform PCA on the mosaic to calculate the eigenvectors (i.e. the principal component spectra shown in Figure \ref{fig:components}) for a particular flyby; the eigenvalues and two primary components for T8 are shown in Figure \ref{fig:PCAimages}. \\



The principal components from any one flyby or VIMS cube can then be used to decompose new VIMS cubes in the same principal component space. The choice of eigenvectors affects the resulting spectrum because the eigenvectors necessarily contain some information specific to a flyby (geometry, illumination, etc). We therefore use three sets of eigenvectors for each flyby to obtain the general behavior of the spectra: eigenvectors from regional maps from T8 (because of its good views of Tui known to exhibit absorption feature and decent SNR for Hotei), T49 (best views of Hotei), and the flyby in question (i.e. the observation dataset itself). \\
 
\begin{figure*}
\includegraphics[width=\textwidth]{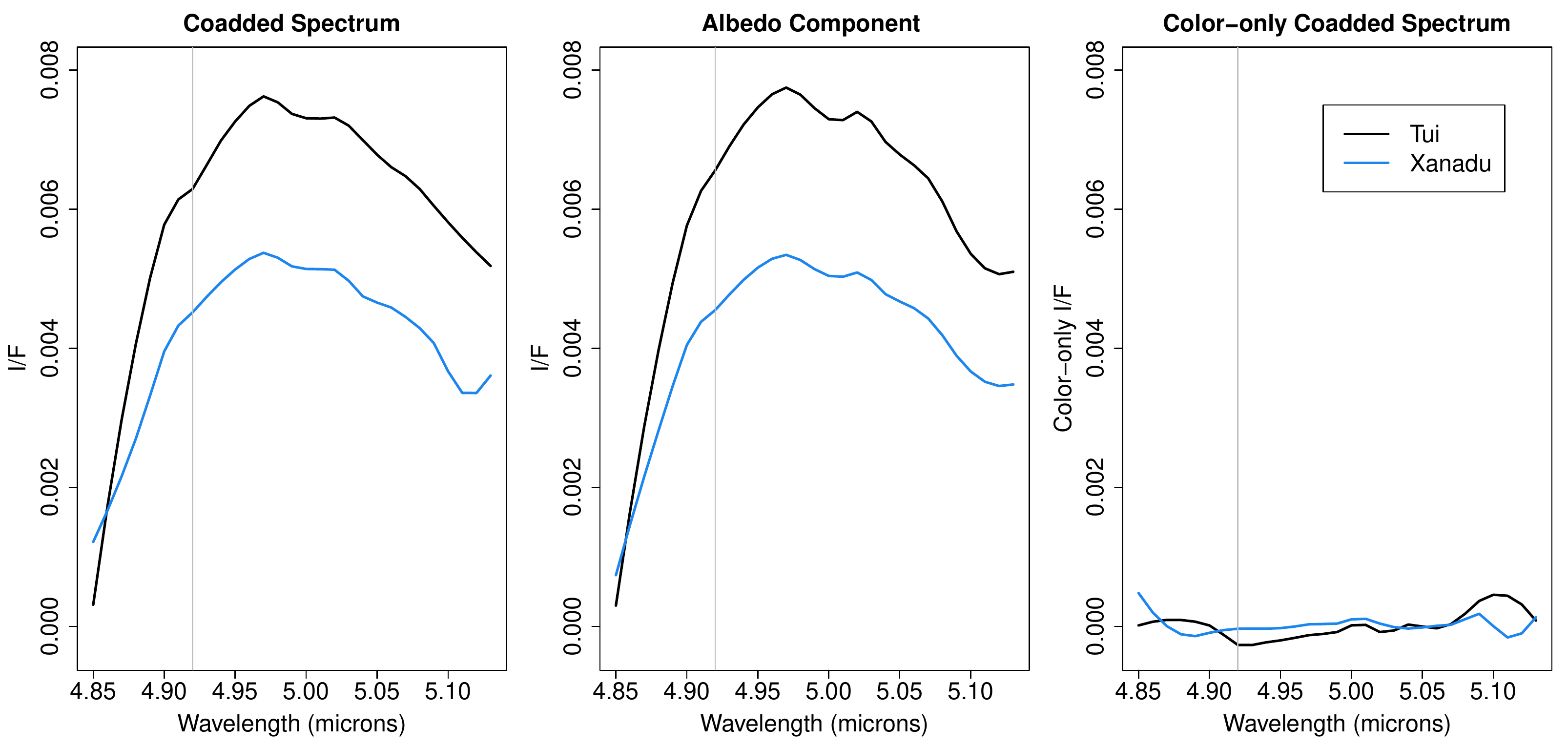}
\caption{Example of subtracting the primary component identified by our principal component analysis from a coadded spectrum to better observe spectral features in the spectra of Tui (black) and Xanadu (blue): (left) raw coadded spectrum from T8, (middle) the albedo component shown in Figure \ref{fig:PCAimages}, and (right) the resulting color-only spectrum. The first component of PCA explains the vast majority of the variance of the spectral data, so when we subtract it from the original spectrum, a fraction of the signal is left. In this color-only spectrum, it is much easier to distinguish differences in the shape of the spectra of Tui and Xanadu.}
\label{fig:PCAsubtract}
\end{figure*}

\subsection{Processing the color-only data}
To obtain the necessary signal-to-noise ratio, we coadd pixels from images within the same flyby, weighting by number of pixels from each image as well as their exposure times, and calculate the standard deviation of the mean of all selected pixels. These pixels come from the color-only images, which is why the corrected I/F values plotted here can be negative--we've subtracted out the common albedo component. Figure \ref{fig:PCAsubtract} demonstrates this effect: the first component explains the largest percentage of the observed spectrum, so subtracting it yields an overall smaller signal. Variances from the spectral mean due to absorptions--that is, chemical composition-- are, however, easier to distinguish in this color-only space. At the far right of Figure \ref{fig:PCAsubtract}, the difference in the spectrum of Tui Regio (black) stands out from the relative flatness of Xanadu's (blue) after the albedo subtraction. \\

 An algorithm reads in the color-only image, selects pixels whose center latitude and longitude are within the feature of interest (e.g. Tui, Hotei, Xanadu, etc.), weights the color-only I/F appropriately, and returns the average spectrum for each flyby. \\

\begin{figure*}
\includegraphics[width=1\textwidth]{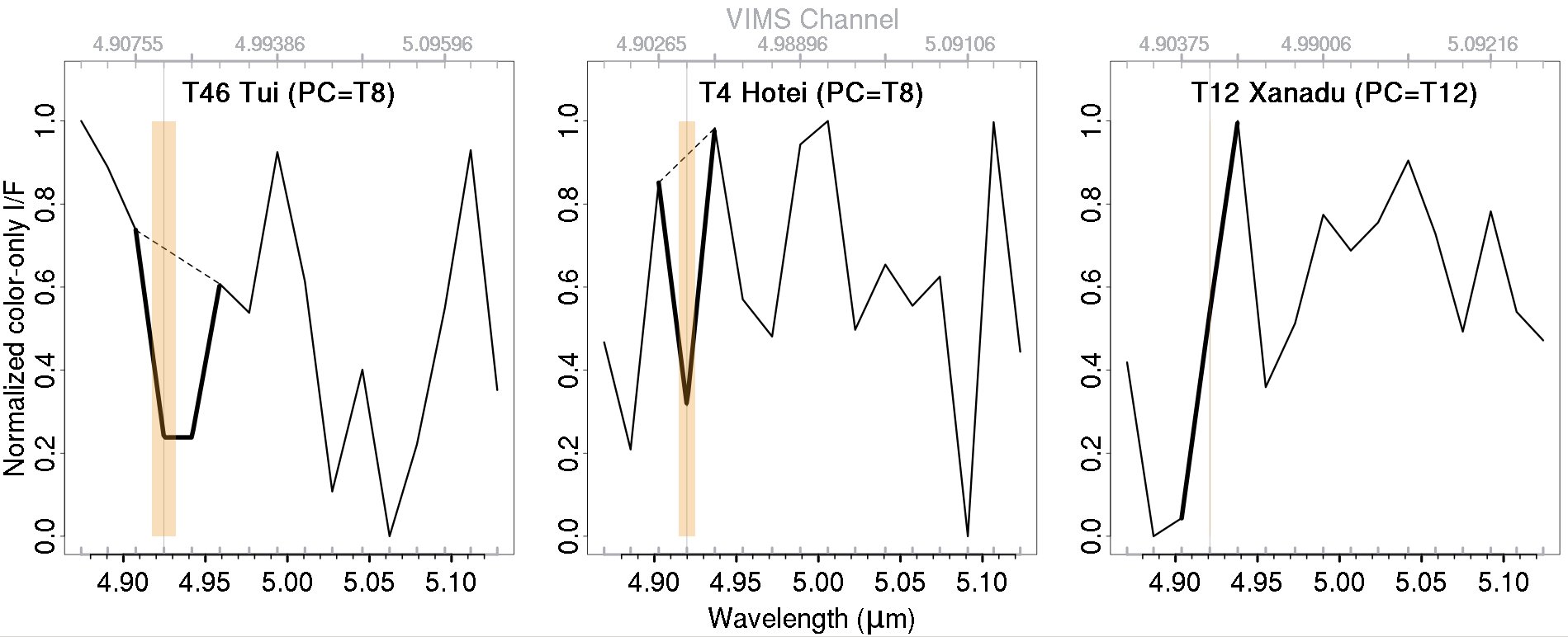}
\caption{Calculating equivalent widths from color-only I/F as a function of wavelength for Tui, Hotei, and Xanadu with data from T12. The thick dashed line represents the continuum estimation while the thick, bolded line represents the analyzed part of the spectrum. These spectra are offset by the value of the spectra at 4.92 $\mu$m and normalized to the maximum.  The vertical grey line is drawn at the VIMS channel closest to 4.92 $\mu$m. The orange box represents the area between the continuum and observed spectra; thus its width is the equivalent width. For Xanadu, which does not have a minimum at or near 4.92 $\mu$m, the equivalent width is zero. The right-most panel represents case (1) where the ``dip" of the 4.92 \um feature is calculated from four spectral points, the center panel is an example of case (2) where it is calculated from three, and the right-most panel is representative of case (3) where there is no feature at 4.92 $\mu$m. }
\label{eqwidthex}
\end{figure*}

We quantify the depth of an absorption feature using the equivalent width metric, the width of a rectangle whose height is unity and whose area is equal to the area between the continuum and the observed spectrum. For Titan's noisy 5 \um surface spectra, we define the continuum to be a straight line drawn between the two endpoints of the 4.92 \um feature. There are three possible cases for identifying the feature, shown in Figure \ref{eqwidthex}: (1) a dip or peak with extrema two VIMS channels wide, (2) a dip or peak with extrema at one VIMS channel, and (3) no peak or dip. For the first and second cases, we define the feature by either three or four points and the third case by two or three (depending on the center criteria, discussed below). We employ this uniformity to reduce the subjectivity of our analysis. \\

We do not know the exact center of the absorption feature at 4.92 \um and the wavelengths sampled by the VIMS channels shift long ward as a function of time\footnote{Documentation of this phenomenon can be found at http://atmos.nmsu.edu/data\_and\_services/atmospheres\_data/Cassini/vims.html}. We thus calculate the equivalent width twice over the entire dataset: (1) identifying the center of the absorption feature at 4.92 $\mu$m and (2) identifying the center at whatever VIMS channel is closest to 4.92 $\mu$m. This distinction only significantly affects flybys in our dataset after T14, where the shift in wavelength from the original channel (4.91983 $\mu$m) becomes greater than  0.0019 $\mu$m ($\sim$ 0.1 spectels). For congruency and ease of comparison between the different flybys, we display the results from method (1) in Figures \ref{fig:depths}-\ref{fig:evapeqwidth}, though the results of the two methods do not differ enough to affect our conclusions. \\

The criteria for a dip in the spectrum to be an absorption are two-fold. First, the equivalent width must be positive; that is, the observed spectrum values at 4.92 \um must be smaller than those of the continuum spectrum. Second, the equivalent width of Tui or Hotei's spectrum  must be larger than that of Xanadu, calculated in the same manner. Xanadu, an equatorial region known for its river networks and mountains, has its own unique spectral unit: bright short of 5 \um with a low 2.7/2.8 ratio \citep{2007mapping}. Because it is not expected to exhibit the absorption at 4.92 $\mu$m, we use Xanadu's equivalent width as our control. Examples of these calculations from T10 are shown in Figure \ref{eqwidthex}. The control region, Xanadu, sometimes has a negative correlation (i.e. a peak at 4.92 $\mu$m) or, as shown in Figure \ref{eqwidthex}, no feature at 4.92 $\mu$m. \\

\section{Results}
\label{Results}
\subsection{Behavior of the Absorption at Tui and Hotei}
\begin{figure}
\begin{center}
\includegraphics[width=0.5\textwidth]{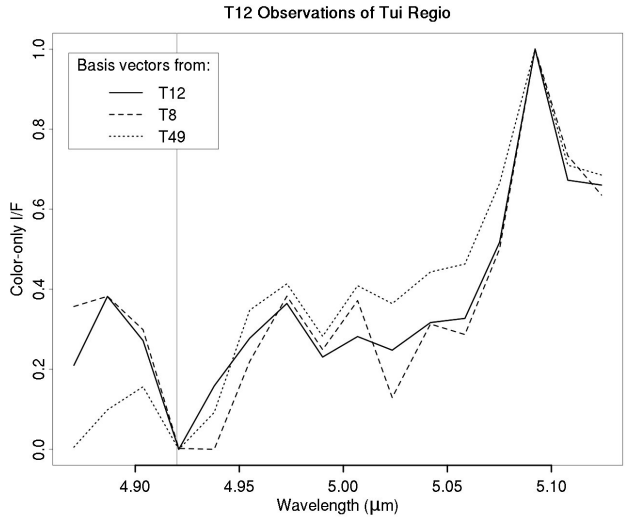} 
\end{center}
\caption{VIMS data for Tui Regio from T12 shown as spectra created by adding all pixels that span Tui Regio in the color-only image using principal components from T12 itself (solid), T8 (dashed), and T49 (dotted). The color-only I/F are normalized to the average coadded I/F of Tui Regio. The grey line of the spectra lies at 4.92. The choice of basis vectors  affects the depth of the absorption feature as the principal components trained from each flyby are not identical.}
\label{fig:Tuispecimages}
\end{figure}

In Figure \ref{fig:Tuispecimages}, we show an example of the color-only 5 \um spectra for Tui Regio using the data from T12. As described in Section \ref{sec:PCA}, the color-only spectrum is independently calculated with the three different sets of principal components generated from different training sets: from the T12 flyby (solid line), from T8 (dashed), and from T49 (dotted). The absorption at 4.92 \um is present in each spectra, though the equivalent width differs with eigenvector to varying degrees. Plotting raw VIMS spectra from different flybys reveals the same problem: the overall amplitude of the 5 \um window (indeed of all Titan's spectral windows) is a complicated function that relies, in part, on viewing geometry. \\

\begin{figure*}
\begin{center}
\includegraphics[width=0.8\textwidth]{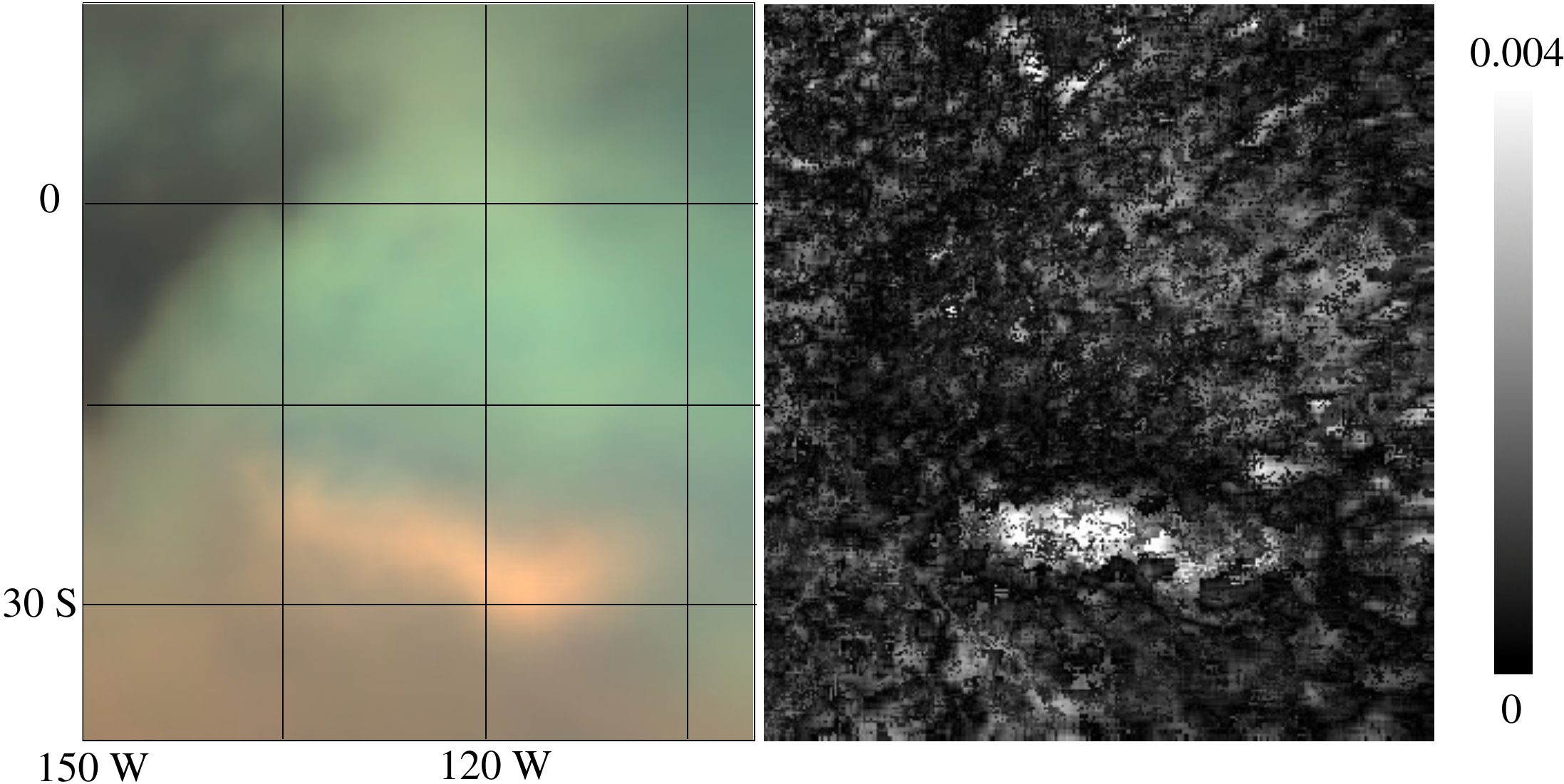}
\end{center}
\caption{Correlation of the spatial extent of Tui Regio and the absorption feature at 4.92 $\mu$m. (left) Tui Regio is the bright pink feature in the VIMS data from Ta, Tb, and T8 shown here as a coadded, cylindrically projected map where north is up (R = 5  $\mu$m, G = 2  $\mu$m , B = 1.3  $\mu$m). Xandau is the green-blue region to the north of Tui. (right) We plot the difference between the equivalent width calculated at 4.92 $\mu$m and the average of the equivalent width of any other dips/peaks in the coadded spectrum of each pixel. Larger difference values (whiter color in the plot) indicate the extent to which the equivalent width of an absorption feature is larger than the effective noise level. }
\label{fig:Tuieqwidthimages}
\end{figure*}

To demonstrate the spatial correlation between the 4.92 $\mu$m absorption feature and the spectrum of Tui Regio, we independently calculate the equivalent width for each pixel of a map of coadded images from a single flyby (rather than creating a coadded spectrum from all pixels within a region as described in Section \ref{sec:PCA}). The signal-to-noise ratio is still too low to observe the absorption feature in a single pixel's spectrum, even for pixels from maps created with multiple images from a single flyby. We therefore run this analysis with spectra coadded from multiple flybys. The map shown in Figure \ref{fig:Tuieqwidthimages} is the coadded spectra of Ta, Tb, and T8 coverage of Tui Regio and Xanadu. These three flybys have similar viewing geometries and exhibit the 4.92 $\mu$m absorption in Tui's spectra. \\

We include the coadded VIMS cylindrical map at the left of Figure \ref{fig:Tuieqwidthimages}. The right image plots the difference between the equivalent width calculated at 4.92 $\mu$m and the average equivalent width of any other dips or peaks in the pixel's spectrum. The larger the difference (whiter values), the more the 4.92 $\mu$m dip is above the pixel's effective noise level, which we estimate as the extent of the second smallest dip or peak of the spectrum.  Comparing the two images reveals that the 5-$\mu$m-bright spectral unit at Tui and the large regions of 4.92 \um absorption spatially correlate. Xanadu, the blue-green region north of Tui, does not demonstrate the same correlation. \\

\begin{figure*}
\includegraphics[width=1.1\textwidth]{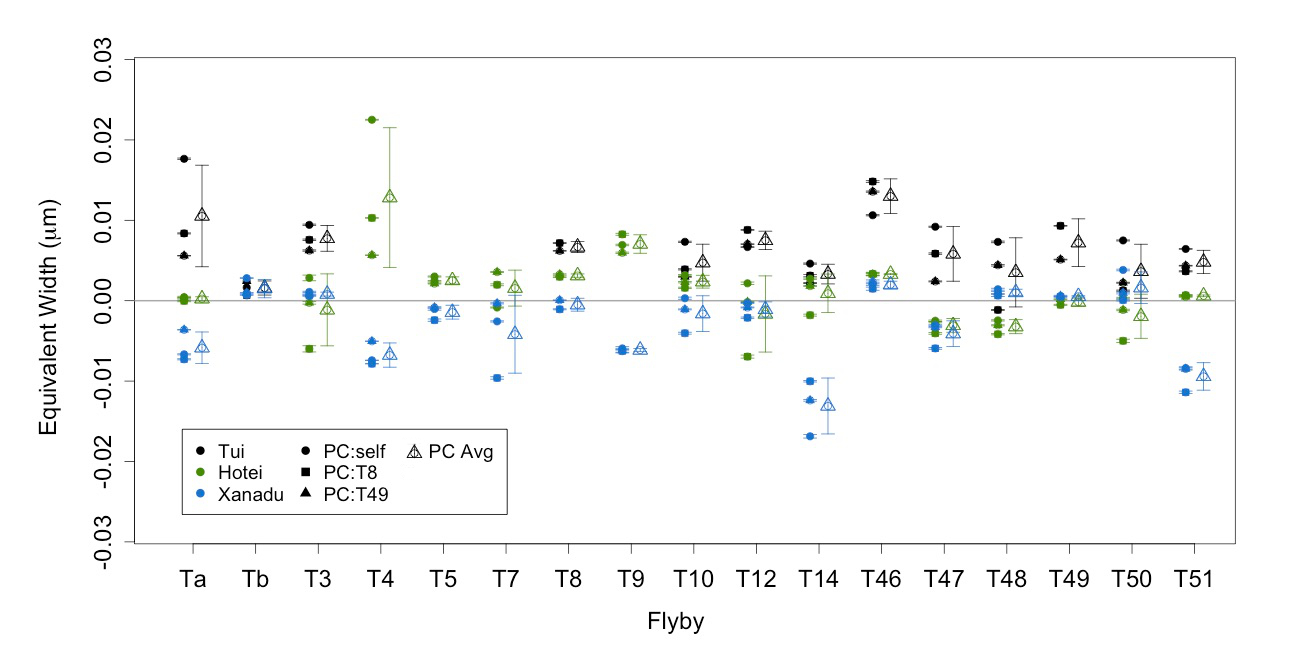}
\caption{Equivalent width of the absorption feature at 4.92 \um as a function of flyby for Tui (black), Hotei (green), and Xanadu (blue). Each shape represents which eigenvector was used to project the data: from the flyby in question (circle), from T8 (square), and from T49 (triangle). The error bars are derived from the standard deviation of the mean for each pixel's color-only I/F at a particular wavelength. For the average from the three equivalent widths, the error bars are derived from the standard deviation of the mean from the three principal component values. }
\label{fig:depths}
\end{figure*}

\begin{figure}
\begin{center}
\begin{tabular}{c}
\includegraphics[width=0.45\textwidth]{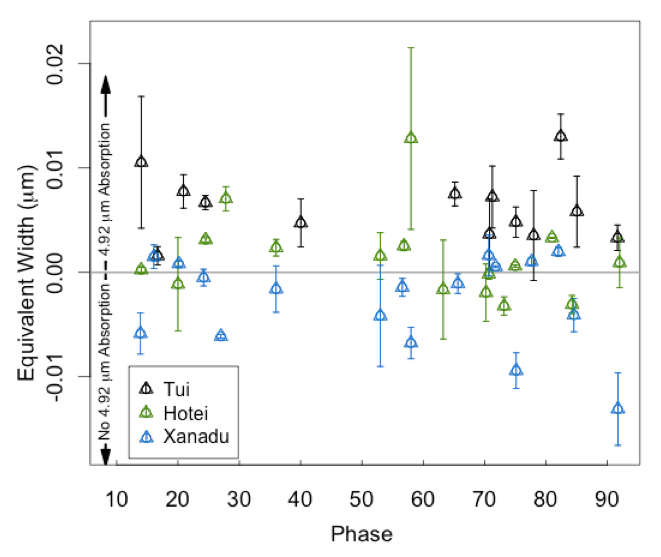}\\
\includegraphics[width=0.45\textwidth]{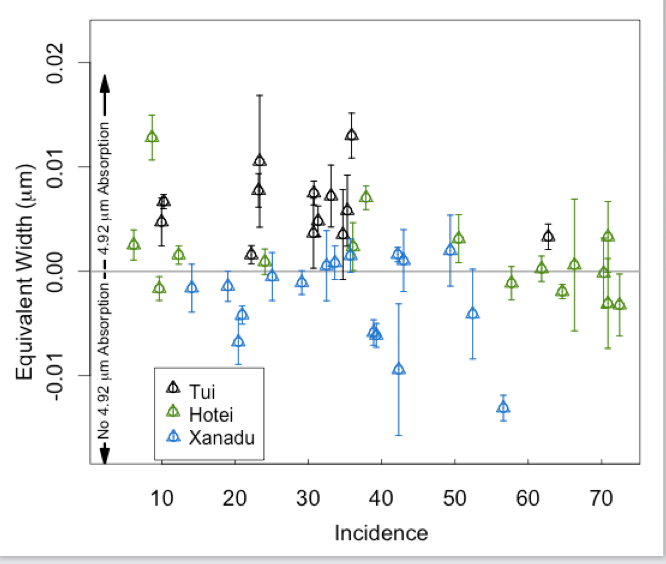}\\
\includegraphics[width=0.45\textwidth]{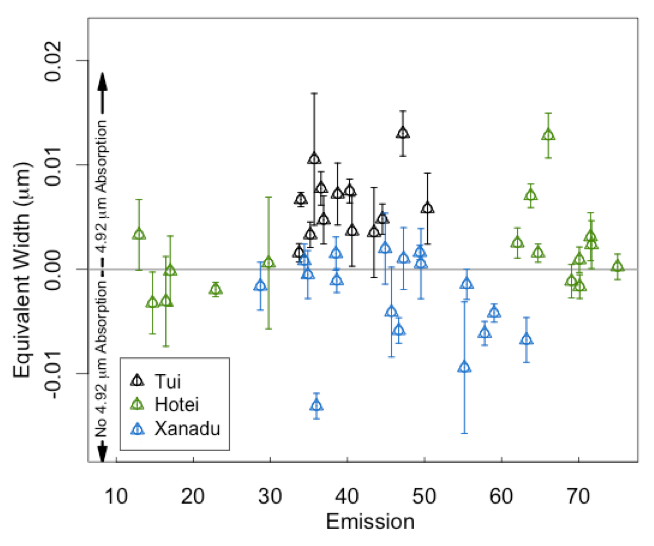}\\
\end{tabular}
\end{center}
\caption{Equivalent width as a function of viewing geometry: phase (top), incidence (middle), emission(bottom) with the same color and shape scheme as Figure \ref{fig:depths}. We only plot the average of the three basis vectors for ease of interpretation; the bars represent the standard deviation from the mean of the equivalent widths calculated with each basis vector. }
\label{fig:eqwidthvangle}
\end{figure}

We show the equivalent width of the 4.92 \um feature at Tui, Hotei, and Xanadu as a function of flyby in Figure \ref{fig:depths} and as a function of phase, incidence, and emission angles in Figure \ref{fig:eqwidthvangle}. The data are colored according to feature (Tui is black, Hotei is green, Xanadu is blue) and grouped by principal component basis vectors (circles are the flyby in question, squares are T8, and triangles are T49). The bars in Figures \ref{fig:depths}-\ref{fig:evapeqwidth} represent the random error arising from the spread in color-only I/F for pixels spanning a surface feature. The standard deviation of the mean calculated from the original selection of pixels is propagated through the calculation of the equivalent widths. \\

The variance in pixel color-only I/F is small for all cases-- often ``error" bars barely clear the size of the point plotted. This indicates that the random error is relatively negligible thanks to the coaddition of pixels in our PCA-correction method. However, this does not take into account the systematic error. If all geometry and viewing effects were accounted for by subtracting the first principal component, then we might expect that differences between the equivalent widths derived from different basis vectors would be insignificant. As they are not, we know that the systematic error is not perfectly accounted for with this method. The average equivalent width between the values calculated with different basis vectors is now shown to the right of the eigenvectors as a circle-filled triangle; the standard deviations of each set are normally distributed with the peak frequency at zero. When comparing the overall behavior of basis vector-derived equivalent widths as a function of flyby, the trends are the same. We thus discuss our results in light of the average behavior and the standard deviation of each eigenvector from this mean. \\

{A linear regression fit through the average equivalent width values for Tui and Hotei reveals that it is reasonable to model the equivalent width as time-independent. The same cannot be said for Xanadu, as we expect for our control feature; the data is randomly distributed around the proposed model. The fit equivalent width is 0.007$\pm$0.002 for Tui. A Bonferroni-adjusted test reveals no outliers, but according to our criteria, Tui shows the absorption feature in all but two (Tb and T50) of the thirteen flybys in which it is observed. In flybys Tb and T50, the calculated equivalent widths for Tui are of the same order as that of Xanadu, our control feature, and are thus too close to definitively call absorption cases. Note that our analysis agrees with that of \citet{McCord2008}, as we observe an absorption feature in the T3 data. \\

Fitting the average equivalent widths for Hotei as independent of flyby gives a value of 0.004$\pm$0.002. As with Tui, no statistical outliers are present, though we exclude six flybys where Hotei's calculated equivalent width is below that of Xanadu's (T3,T12,T47,T48,T49,T50). Thus,} Hotei only shows the absorption feature in eight of its sixteen observations (T4,T5, T7,T8, T9, T10, T14, and T46), with two flybys significantly above the time-independent average (T4 and T9). Two flybys have equivalent widths of near zero (Ta and T51).\\

When plotted as a function of viewing geometry (Figure \ref{fig:eqwidthvangle}), we do not observe any obvious trend to explain why the absorption feature is different from the time-independent average in some flybys and not others. Comparing each viewing angle with equivalent width via a Kendall tau test reveals no correlation. (With coefficients $\leq$ 0.1 for each, we cannot reject the null hypothesis.) For example, the two flybys during which the equivalent width for Tui Regio is significantly above the average, Ta and T46, are both near specular, but otherwise not similar in geometry. Ta is low phase with medium incidence and emission while T46 is high phase with medium incidence and emission. Furthermore, the flyby during which Tui's equivalent width is significantly below the average, T14, has almost identical viewing geometry to T46. Hotei's significant cases, Ta and T51 below the average and T4 and T9 above, are also near specular. But, as with Tui Regio, T51 and T4 are high phase while Ta and T9 are low phase. Therefore, our results do not support that any one value of phase, incidence, emission either independently or cumulatively is responsible for controlling equivalent width. We discuss possible explanations for this behavior and future analyses in Section \ref{Discussion}. \\

\subsection{Evaporites}
\begin{figure*}
\includegraphics[width=1\textwidth]{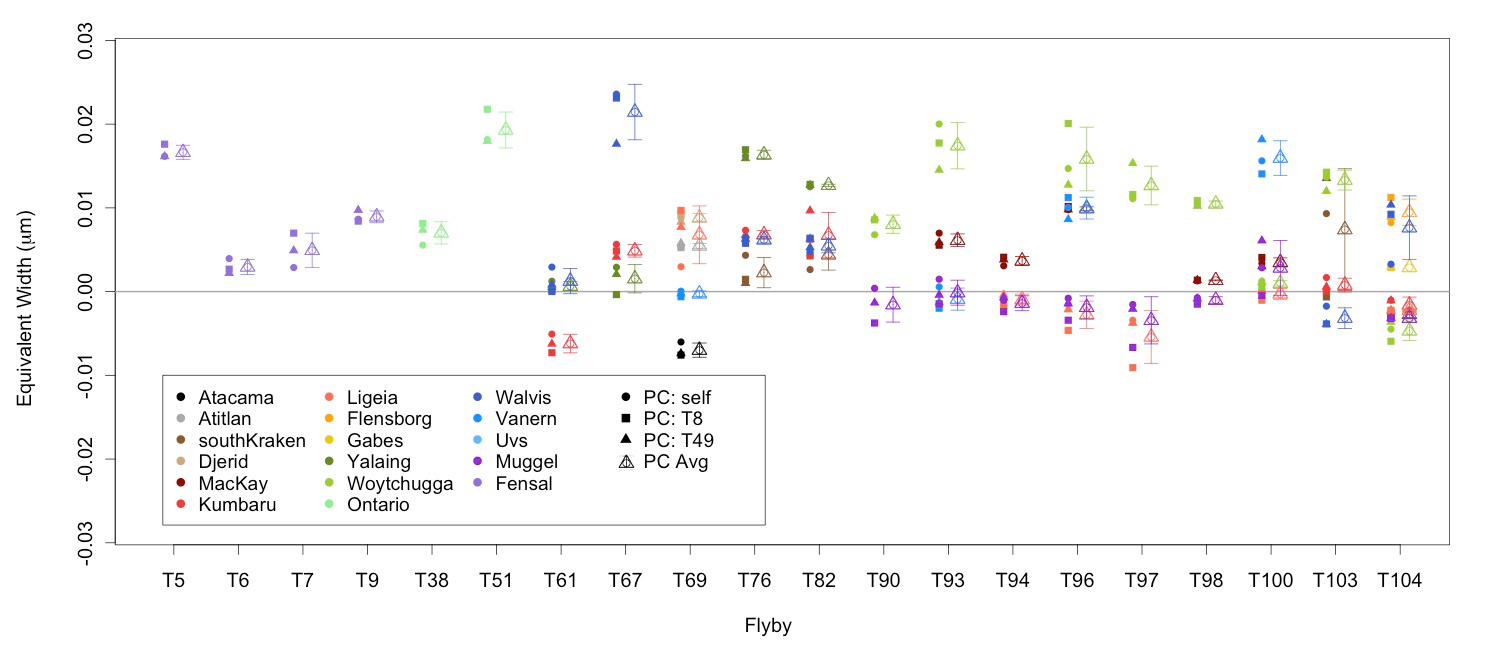}
\caption{ Equivalent width as a function of flyby for evaporites. Shapes correspond to which flyby the eigenvector for the PCA was derived from: the flyby in question (circle), T8 (square), or T49 (triangle). The error bars for each particular principal component are not shown as they are smaller than the point size. For the average from the three equivalent widths, the error bars are derived from the standard deviation of the mean from the three principal component values.}
\label{fig:evapeqwidth}
\end{figure*}

\begin{figure*}
\includegraphics[width=1.8\columnwidth]{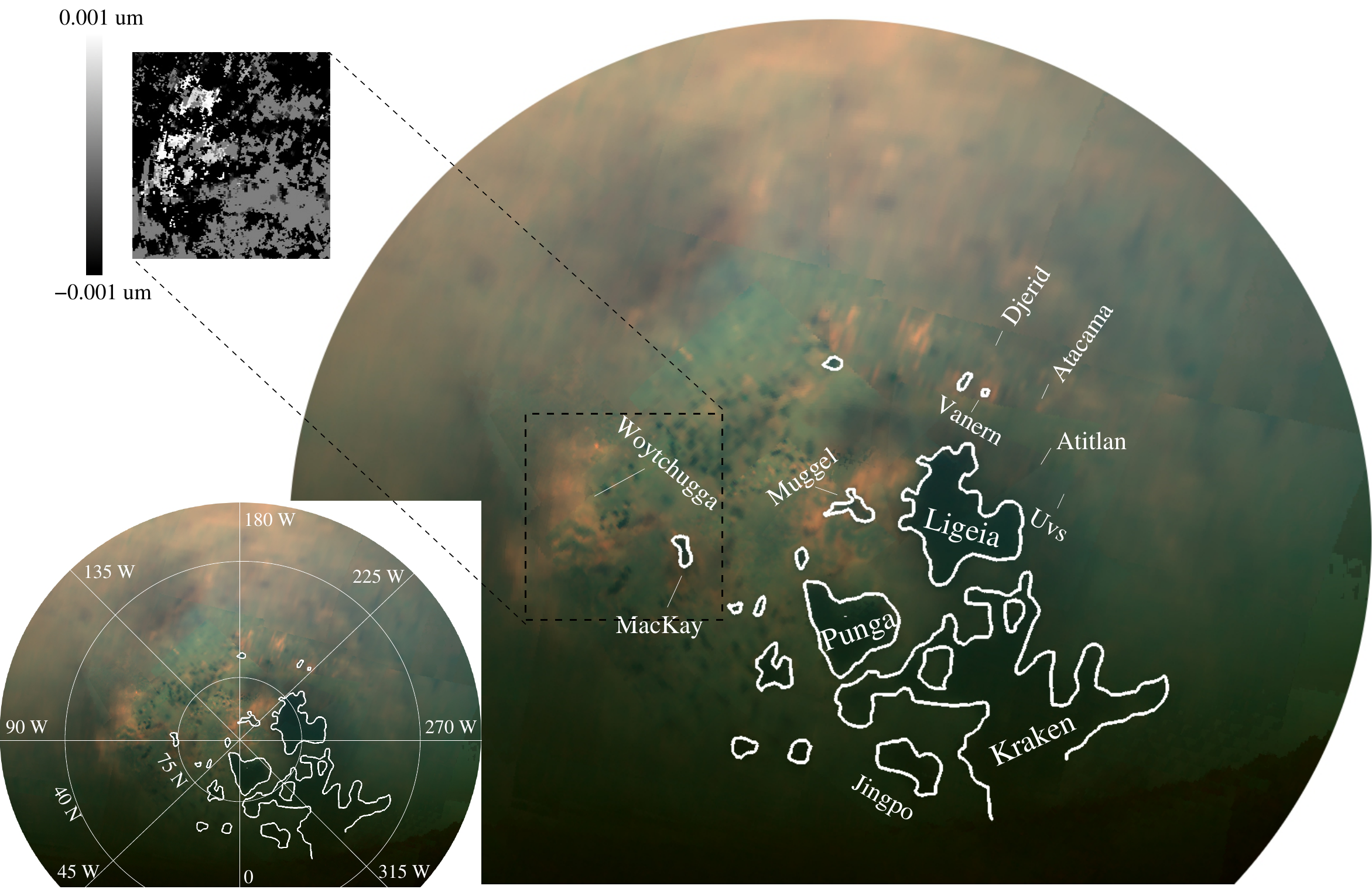}
\caption{ Orthographic projection of Titan's north pole in VIMS data (T93, T94, T96, and T97) with color scheme R = 5 $\mu$m, G = 2 $\mu$m, and B = 1.3 $\mu$m. The inset is the difference between the equivalent width at each pixel and the effective noise level of a coadded image of Woytchugga Lacuna and MacKay Lacus (created with coadded data from T96, T97, and T98), similar to the right panel of Figure \ref{fig:Tuieqwidthimages}. The extent of 5-$\mu$m-bright Woytchugga spatially correlates with positive equivalent widths at 4.92 $\mu$m well above the noise level (white values of the inset). We expect that the patches of dark within Woytchugga are a sampling effect. }
\label{fig:nportho}
\end{figure*}

We plot equivalent width as a function of flyby for evaporites in Figure \ref{fig:evapeqwidth} with the same shape scheme as we used in Figure \ref{fig:depths}. As it is difficult to see the most northern features in the cylindrical projection of Figure \ref{fig:global}, we show in Figure \ref{fig:nportho} a labeled orthographic projection of the best VIMS data (T93-T97) of Titan's north pole in the same color scheme as Figure \ref{fig:global}. The inset of Figure \ref{fig:nportho} shows the equivalent width-to-noise calculation of Figure \ref{fig:Tuieqwidthimages} for Woytchugga Lacuna and MacKay Lacus\footnote{\emph{Lacuna} refers to a dried or potentially ephemerally filled lakebed while a \emph{lacus} is a filled lake.}. (There are enough data from T96-T98 of similar viewing geometry for these two features to construct images of the necessary SNR per pixel.) Woytchugga, whose positive equivalent widths at 4.92 $\mu$m stand out above the noise level as an absorption feature, appears white in the inset. The region surrounding MacKay Lacus, which is also 5-$\mu$m-bright, also shows an absorption feature when coadding pixels from across the feature, but individually, no one pixel is above the noise level. Thus, it is indistinguishable in the inset.\\

It is less useful to conduct the linear regression analysis described above for Tui and Hotei with these evaporite cases, as most do not have overlapping error bars and have only a handful of observations. It is interesting to note, however, that the spread between values derived from different basis vectors is much smaller for the evaporites than for Tui, Hotei, and Xanadu. The evaporites with the largest spread are those with the largest surface area-- Woytchugga, Flensborg Sinus, and south of Kraken. Thus, we interpret that another source of unaccounted error is the spectral blending unavoidable at km/pixel resolution. For example, in the higher resolution data of Hotei in T48 and T49, \citet{2009Icar..204..610S} identify patches of dark blue spectral units. The polygons used to define the extent of Hotei for our pixel selection was drawn on these higher resolution maps, but this cannot fix coarser resolution data.   \\

According to the calculated equivalent widths shown in Figure \ref{fig:evapeqwidth}, we can group our results into three groups: evaporites that show the absorption feature in all observations, in most observations, and in at most one observation. There are eight evaporite deposits that demonstrate the absorption feature in every observation: deposits located in west Fensal (4 flybys); Ontario Lacus's shoreline deposits (2); Kraken Mare's Flensborg Sinus\footnote{\emph{Sinus} is the International Astronomical Union designation for bays on Titan.} and Gabes Sinus (both only observed once); Djerid Lacuna (1); Atitlan Lacus (1); Uvs Lacus (1); and MacKay Lacus (5). \\

In all but one of observation, Woytchugga Lacuna (7/8 flybys) and Walvis Sinus (5/6) demonstrate the 4.92 \um absorption feature, while Kumbaru Sinus (4/6) shows it in all but two observations. The 5-$\mu$m-bright deposit at the north end of Yalaing Terra and the deposits south of Kraken Mare both demonstrate absorption features at 4.92 \um in two out of the four observations of each feature. Whether these differences between flyby are due to viewing geometry, surface roughness, evaporite composition, etc. cannot be addressed by the methods used here and is thus beyond the scope of this paper. It is instead further evidence for the complexity of the problem evident in the results for Tui and Hotei. There are also evaporites that do not exhibit the absorption feature in a majority of their observations: Muggel Lacus, Vanern Lacus, Ligeia Mare shoreline deposits, and Atacama Lacuna. \\

\section{Discussion and Conclusions}
\label{Discussion}
Not all  5-$\mu$m-bright terrains demonstrate the absorption feature at 4.92 $\mu$m, nor do those that do show the absorption show it in all observations. Our results cannot detail what physical characteristics set this behavior (i.e. relative abundances, micro-scale surface structure, etc.). To do so, more rigorous modeling of each feature's phase function will require a full radiative transfer correction for the atmosphere, our next project. However, our data, if true, do illuminate bulk spectral behavior, which we discuss below: what factors probably play in our results for Tui, why Hotei is different, and the heterogeneity of evaporite deposits.\\

\subsection{Tui Regio}
The equivalent widths of the 4.92 $\mu$m absorption feature in the spectra of Tui Regio are probably a function of viewing geometry, exposure time, the VIMS pipeline flat-field calibration, particle size, scattering properties of the grains of absorbent material, as well as the optical properties of surrounding materials. Generally, mid incidence angle observations yield a ``stronger" absorption for Tui Regio, though if our estimated errors are robust, there is little difference between the extent of the absorption feature between flybys. \\

Titan evaporites could be crystalline like Earth evaporites as the end result of their similar formation process; such a structure would help explain why the 5-$\mu$m-bright material is bright at all wavelengths. As such, evaporites could have a preferred viewing geometry at which the absorption feature could appear stronger due to increased internal scattering before the light refracts out of the crystal and returns to the detector. Such an effect would be complicated by surrounding material and different particle sizes. \\

\subsection{Differences between Hotei and Tui}
The equivalent widths of the 4.92 \um absorption at Hotei Regio are generally smaller than those at Tui for the same flyby. Interestingly, Hotei does not show the same angular dependence as Tui: Hotei's absorption features are wider at smaller incidence and larger emission angles. However, the averages for both features in the flyby analysis are close enough to be within the estimated error of the fit, what we might expect if these features contain the same material responsible for the 4.92 $\mu$m absorption. \\ 

If the material responsible for this absorption is present at both Tui and Hotei, why might it demonstrate different spectral behavior in flybys where we observe both features? Our analysis of the influence of viewing geometry is not sufficient to explain this behavior-- there must be other driving factors. For example, it has been shown by \citet{1999Icar..137..235S} that absorption depths for regolith-like surface material are strongly dependent on particle size; larger particles yielded a stronger absorption for the lunar soils of that study. Thus, it could be that the 4.92 $\mu$m absorbing grains are of different sizes at Tui Regio and Hotei Regio. Or, if Hotei's 4.92 $\mu$m absorbent material is covered by, say, larger grains not present or not of the same size than at Tui, the depth of the absorption feature might be dampened in different viewing geometries (i.e. flybys) for the different basins. Work by \citet{2015Icar..250..188P} demonstrated that the overall phase curve behavior is largely controlled by the brightest, most abundant, and least isotropically scattering particles in a material. Of course it also could be that the albedo subtraction doesn't work as seamlessly as we have assumed, but the results for the evaporite cases demonstrate that material variability is not an unreasonable explanation. \\

\subsection{Evaporites}
What distinguishes evaporite deposits that do show the 4.92 \um absorption feature from those that don't? There is no simple, common geomorphological characteristic. Woytchugga Lacuna is a large (66,700 km$^2$), completely dry lakebed. Ontario Lacus is a lake with exposed evaporite along its eastern shorelines. MacKay Lacus is a partially-filled lake at $\sim$ 75$^{\circ}$ N.
Vanern Lacus is a 400 km$^2$ partially-filled lake south of Ligeia Mare with a bathtub ring of evaporite and Djerid Lacuna is a nearby dry lakebed of similar size. Yet all of these features demonstrate an absorption at 4.92 $\mu$m in most of their observations. \\

\begin{figure}
\includegraphics[width=\columnwidth]{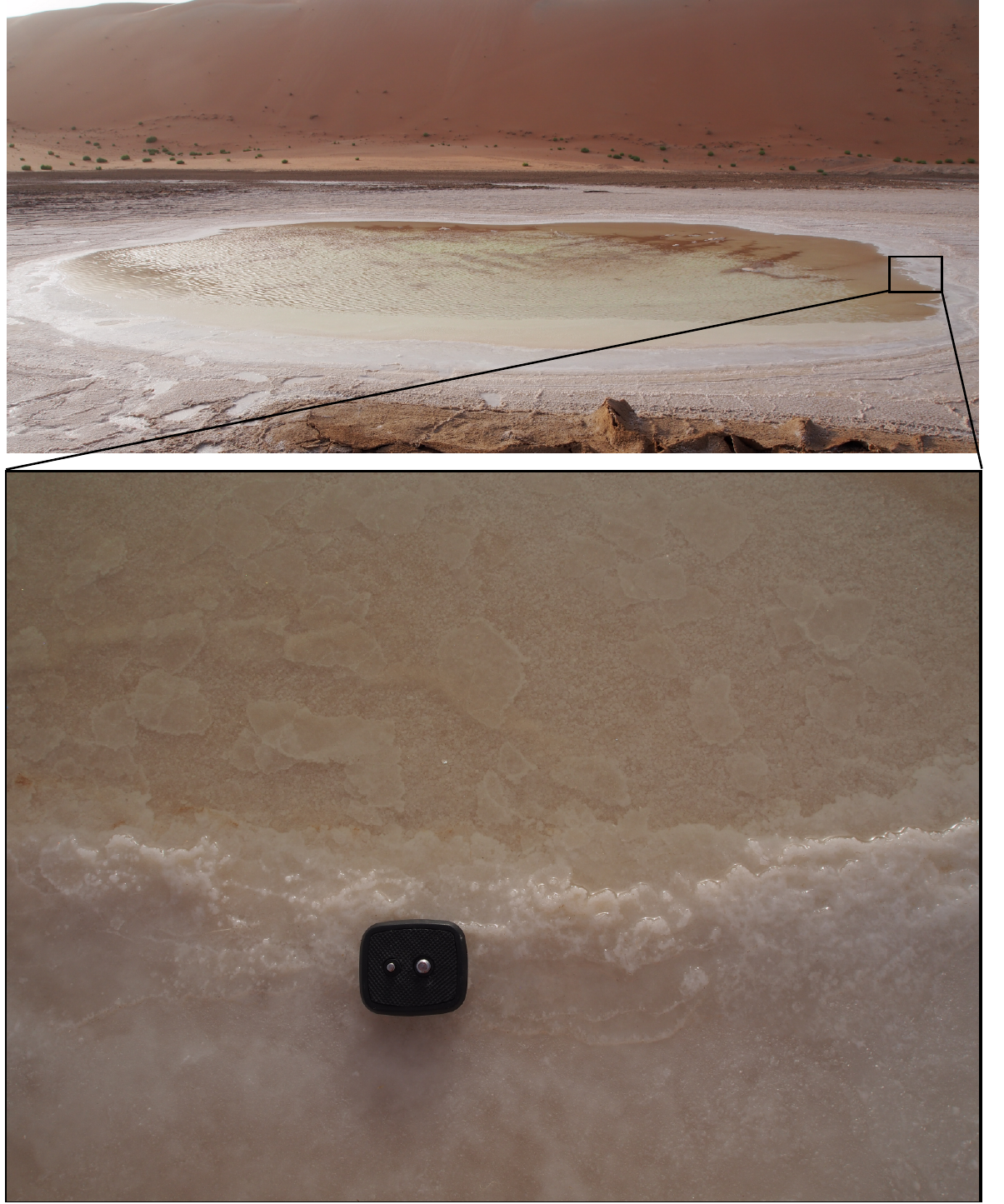}
\caption{Earth evaporites forming in an interdune at Liwa, UAE. The extent of the liquid-filled sabkha is shown at top. In the inset, gypsum evaporite falls out as thin sheets at the shoreline while the bottom of the pond is covered by cubic halite structures. The camera mount included for scale is 5.5 cm across. }
\label{uaeevap}
\end{figure}

Recent work by \citet{Cordier2015} shows that for solutions with several solutes (each of different solubility in a Titan liquid), the evaporites will precipitate out in layers based on their solubilities; compounds with the largest solubilities will stay in solution the longest. In Figure \ref{uaeevap} we show an example of this process on Earth in a wet interdune in Liwa, UAE (top photo). The inset shows thin sheets of gypsum evaporites along the edges of the receding shoreline while cubed halite is at the bottom of the liquid. This separation is due to the respective solubilities of the two compounds as well as the saturation of the solution.\\

\textbf{Thus, it could be that} each of these evaporite locations that do show the absorption are far enough along in the drying-out stage to have precipitated out the more soluble 4.92 \um absorbent material. Djerid and Woytchugga are at the end point of drying out (i.e. there is not enough liquid left to be detectable by VIMS).While, based on comparing the areas of liquid-covered surface and evaporite covered surface, Vanern and Ontario have receded by 90\%  and 15\% their original areas, respectively. If our results are indicating that the evaporite deposits demonstrating the 4.92 \um absorption feature have reached some endpoint where the most soluble material has precipitated out, then the work of \citet{Cordier2015} would point to butane and acetylene as candidates for the material responsible for the 4.92 \um absorption.  \\

Of course it could also be that the non-absorption cases (Muggel Lacus, Ligeia Mare shorelines, and Atacama Lacuna) have a different solution composition. At 83$^{\circ}$ N, Muggel is nearer to the north pole than the other deposits considered here. \citet{2014GeoRL..41.5764L} discuss how more frequent rains at the northern-most parts of the pole might be responsible for a ``salinity" dichotomy between the more northern Ligeia Mare and more southern Kraken Mare. Perhaps, then, Muggel and MacKay only precipitate out the least soluble sediments (and thus those that do not exhibit the 4.92 \um absorption), as before the 4.92-$\mu$m-absorbent material falls out of solution, a new influx of methane halts evaporite formation. Or it could be that the Ligeia watershed washes different sediments or even different amounts of sediments into the sea than Kraken's.\\

As for the less straightforward evaporite deposits that behave more like Hotei (Walvis Sinus, Kumbaru Sinus, Flensborg Sinus, south of Kraken deposits, Yalaing, and deposits located in west Fensal), it is again difficult to pull out viewing geometry dependence of extent of absorption feature depth without more rigorous modeling. Indeed, there are some inconsistencies in the cases we consider as demonstrating the absorption feature, like Woytchugga Lacuna which does not show an appreciable equivalent width in the most recent observations (T104). While we attribute these ambiguities to deficiencies in our current method, we cannot exclusively rule out the possibility that the 4.92 $\mu$m absorption in the studied evaporites might be lost in noise or highly variable due to dynamic surface processes. \\

For example, as documented by \citet{PrecipSurfaceBrightening}, there are VIMS observations of north Yalaing before, during, and after a wetting event. During T61-T67, Yalaing appears 5-$\mu$m-bright, but does not consistently show the absorption feature (note the disagreement between the different eigenvector samples in T67). In VIMS data from T76, during the wetting event, Yalaing Terra appears brighter at 2 $\mu$m, is no longer 5-$\mu$m-bright, and exhibits the absorption. With a thin enough layer, VIMS could still be sampling surface material beneath the 2-$\mu$m-bright covering. Then, after the wetting event (T82) the deposit returns to the 5-$\mu$m-bright spectral unit and, in our results, shows the absorption feature, though to a lesser extent than that observed in T76. \citet{PrecipSurfaceBrightening} propose that the spectral change was probably due to the presence of a transient layer: either volatile frost that sublimes away or a deposition of fine-grained particles that blow away to reveal the original material underneath. Sintering driven by the first scenario could create larger grain sizes that might thus explain the increased absorption feature at flybys during and after the wetting event. Or, it could be that this wetting event instigated a removal process of some surrounding material that promoted detection of the 4.92 \um feature.\\ 

Other evaporite candidates exhibit different values for equivalent width depending on flyby. While it is not clear that our analysis has removed all sources of systematic error, it is also a possibility that these other cases have experienced some change similar to that observed at Yalaing Terra. Unfortunately, there is no evidence in the ISS or VIMS cloud coverage maps to support such an explanation for the dramatic changes in equivalent width for Fensal \citep{2009Natur.459..678R,2011Icar..216...89R,2011GeoRL..38.3203T}. Clouds have frequently been observed at the higher latitudes where the north polar evaporites are, but VIMS data provides no benchmarks for what the surface looked like beforehand. Surprisingly, there has been a relatively few clouds at the north pole in our latest observations, despite global circulation model predictions of a pick up in seasonal weather activity \citep[e.g.][]{2006Sci...311..201R, 2012ApJ...756L..26M}. Thus, there are no observed weather events able to explain, for example, the change between Vanern observed in T93 and T100. \\

Evaporites probably form on larger timescales than the lifetime of the Cassini mission, however. Thus, it is useful to also consider the results of general circulation models, which, for Titan, largely predict greater rainfall at the poles and a relative dearth at the equator \citep{2006Sci...311..201R,2008JGRE..113.8015M}. That is not to say that it is impossible for rain to occur there-- as demonstrated by the VIMS observations detailed above by \citep{PrecipSurfaceBrightening} and ISS observations \citep{2011Sci...331.1414T}. Some GCM results suggest that perhaps this isn't as much of a problem for the evaporites found away from Titan's poles as previously thought. For example, in the model of \citet{2015Icar..250..516L}, pockets of equatorial and midlatitudes experienced surface liquid and drying periods (the conditions necessary for evaporite formation). This model's initial conditions included a 4 m deep methane reservoir with deeper, localized areas to represent the seas and Ontario Lacus. Interestingly, the evolution of such a Titan results in areas of surface liquid change in locations where we observe evaporites (namely, at Woytchugga Lacuna, Tui Regio, Hotei Regio, and Fensal; see Figure 14 of \citet{2015Icar..250..516L}).   \\

\subsection{Conclusions}
We report the observation of the same 4.92 \um absorption described by \citet{McCord2008} in the spectrum of Tui Regio in later observations of that surface feature, as well as in some observations of Hotei Regio, another equatorial basin covered in 5-$\mu$m-bright material. Though we explore the dependence of this absorption feature on viewing geometry, we find that the phase function will require full radiative transfer treatment to solve completely. We also look for the 4.92 \um absorption in the spectra of several evaporites identified by \citet{evaporite,MacKenzie14} and find that while some do, not all evaporites have the spectral feature. We propose that this variance could be due to differences in solution composition, different states of drying, or different surface roughnesses. Our analysis is not able to definitively discern between which of these possibilities could best explain the variable behavior of the 4.92 \um absorption feature.  \\

\acknowledgments
This work was supported by NASA Headquarters under the NASA Earth and Space Science Fellowship Program- Grant NNX14AO30H to SMM and the NASA Cassini Data Analysts and Participating Scientists (CDAPS) Grant NNX12AC28G to JWB. We gratefully acknowledge Sebastien Rodriguez for the helpful comments in his review.

\bibliographystyle{apj}


\end{document}